\documentclass[twocolumn,showpacs,amsmath,amssymb,showkeys,floatfix,prl]{revtex4}
\usepackage{amsfonts,amsmath}
\usepackage{graphicx}
\usepackage{dcolumn}
\usepackage{bm}

\begin{document}

\title{Analysis of three-neutrino oscillations in the full mixing angle space}

\author{D. C. Latimer and D. J. Ernst}

\affiliation{Department of Physics and Astronomy, Vanderbilt University, 
Nashville, Tennessee 37235, USA}
\date{\today}

\begin{abstract}
We use a model, with no CP violation, of the 
world's neutrino oscillation data, excluding the LSND experiments, and search the full 
parameter 
space  ($0\leq\theta_{12} \le \pi/2$; $-\pi/2 \le \theta_{13} \le \pi/2$; and 
$0\leq\theta_{23} \le \pi /2$) for the best fit values of the mixing angles and mass-squared 
differences. We find that the mixing angle $\theta_{13}$ is bounded by $-0.15<\theta_{13}<0.20$
with an absolute minimum at $\theta_{13}=0.12$ and a local minimum at $-0.04$. The importance 
of the negative $\theta_{13}$ region and this structure
in the chi-square space has heretofore been overlooked because the factorization approximation
commonly employed yields oscillation probabilities that are a function of $\sin^2\theta_{13}$.
\end{abstract}



\maketitle

The observation of neutrino oscillations requires a fundamental modification
of the electroweak theory. The simplest, but not totally consistent, method for
accommodating neutrino oscillations into the theory is to 
introduce {\it a posteriori} a
mass matrix and unitary mixing matrix. The standard \cite{PDG} representation of 
the three neutrino mixing matrix is 
\begin{widetext}
\begin{equation}
U_{\alpha k} \rightarrow \left(  
\begin{array}{ccc}
c_{12} c_{13} & s_{12} c_{13} & s_{13} e^{-i\delta}\\
-s_{12}c_{23} - c_{12} s_{23} s_{13} e^{i\delta} & 
c_{12} c_{23} - s_{12} s_{23} s_{13} e^{i\delta} & 
s_{23}c_{13} \\
 s_{12}s_{23} - c_{12} c_{23} s_{13} e^{i \delta} & 
-c_{12} s_{23} - s_{12} c_{23} s_{13} e^{i \delta} & c_{23}c_{13}
\end{array}
\right)\,,
\end{equation}
\end{widetext}
where $c_{\alpha k} = \cos{\theta_{\alpha k}}$,
$s_{\alpha k}=\sin{\theta_{\alpha k}}$,  and $\delta$ is the CP violating phase with
$\theta_{\alpha k}$ and $\delta$ real. We order the mass 
eigenstates by increasing mass, and the flavor eigenstates are 
ordered electron, mu, tau. The bounds on the
mixing angles $\theta_{jk}$ are $0\leq \theta_{jk} \le \pi/2$ and $0\leq \delta
< 2\pi$. In the absence of CP violation, the range of the mixing angles \cite{gluza} is 
$0\leq \theta_{jk} \le \pi/2$ with $\delta=0$ {\it and} $\delta=\pi$; or equivalently 
\cite{angles} take {\it only} $\delta=0$  with $0 \le \theta_{12} \le \pi/2$,
$-\pi/2 \le \theta_{13} \le \pi/2$, and $0 \le \theta_{23} \le \pi/2$. 
Experiments find that $\theta_{13}$ is near zero. The second option 
produces one contiguous allowed region in the parameter space; the former gives two 
disconnected regions for the allowed parameters. In 
particular, oscillation probabilities for $\theta_{13}$ negative are not related to those 
for $\theta_{13}$ positive. 
Parameterization of oscillation solutions by $\sin^2 \theta_{13}$ is thus inadequate.

In vacuo, the probability that a neutrino with energy $E$ and flavor $\alpha$ 
will be detected a distance $L$ away as a neutrino of flavor $\beta$
is given by
\begin{eqnarray}
{\cal P}_{\alpha \beta}(L/E) &=& \delta_{\alpha \beta} \nonumber\\  
&&-4\sum_{\genfrac{}{}{0pt}{}{j,k=1}{j< k}}^3
U_{\alpha j} U_{\beta j} U_{\alpha k} U_{\beta k} \sin^2 \phi^{osc}_{jk}\,,
\label{pee}
\end{eqnarray}
in which $\phi^{osc}_{jk}=1.27 \Delta m_{jk}^2 L/E$ with $L/E$ 
expressed in units of m/MeV and $\Delta m_{jk}^2 \equiv
m_j^2 - m_k^2$ in units of eV$^2$. 
This probability is then to be integrated over the energy spectrum of the neutrinos for 
each experiment. 
 
We construct a model of the data and then, within the model, look for best fit oscillation 
parameters throughout the full range of permitted mixing angles.
Included in the model are data for neutrinos from the sun 
\cite{homestake,gallium,sno,snosalt}, reactor neutrinos
\cite{chooz,kamland}, atmospheric neutrinos \cite{atm}, and beam-stop neutrinos
\cite{k2k}. We, like others, omit from the analysis the LSND \cite{LSND} and Karmen 
\cite{karmen} experiments. 

Experiments for solar neutrinos \cite{homestake,gallium,sno} historically 
measured the survival probability of electron neutrinos, ${\cal P}_{ee}$. Recent 
experiments \cite{sno,snosalt} measure two different neutrino interactions which then 
allow the extraction of ${\cal P}_{ee}$ and the total solar neutrino flux. The measured 
total is in agreement with the theoretical predictions of the standard solar model
\cite{bp2000}. We here use the standard solar model for the production of neutrinos in 
the sun. Each detector measuring solar neutrinos has a different acceptance and 
thus measures different energy neutrinos. 
In order to reproduce the energy dependence of the 
survival rate of electron neutrinos arriving at the Earth as 
seen in the experiments, we invoke the 
MSW effect \cite{msw}. The MSW effect arises because the neutrinos created in 
the sun propagate through a medium with a significant electron density. The 
forward coherent elastic neutrino-electron scattering produces an effective 
change, relative to the mu and tau neutrino, in the mass of the electron
neutrino given by 
$A(r)=\sqrt{2} \,G \,E \,\rho(r)/m_n$, with $\rho(r)$ the electron density at a 
radius
$r$, $G$ the weak coupling constant, and $m_n$ the nucleon mass. In the flavor basis, the Hamiltonian 
then becomes 
\begin{equation}
H_{mat} = U {\cal M} U^\dagger + A(r) \,,
\end{equation}
with ${\cal M}$ the (diagonal) mass-squared matrix in the mass eigenstate basis 
and $A$ the 
$3\times 3$ matrix with the interaction $A(r)$ as the  
electron-electron matrix element and zeroes elsewhere. By diagonalizing this
Hamiltonian with a unitary transformation $D_{\alpha k}(r,E)$, we define local 
masses and eigenstates as a function or $r$ and $E$. Care must be taken so that
$D_{\alpha k}(r,E)$ becomes $U_{\alpha k}$ in the limit of zero electron density.
In the adiabatic limit, which we use, the electron survival probability is 
\begin{equation}
P_{ee}^{ad}(r,E) = \sum_{k=1}^3 D(r,E)_{ek}^2 \,U_{ek}^2\,.
\end{equation}

Neutrinos are produced throughout the sun by various reactions, each with its 
own energy spectrum. The surviving neutrinos are then detected by detectors 
which have a different acceptance for each energy of the neutrino. 
The survival probability for an electron neutrino in a particular 
experiment is given by
\begin{equation} 
P_{ee}^{ex}= \sum_{j=1}^N p_j^{ex} \int_0^{R_\odot}f_j(r) \,dr
\int_{E_{thresh}}^\infty g_j(E) \,
P_{ee}^{ad}(r,E)\,dE\,.
\end{equation}
Here, $j$ labels a particular nuclear reaction; we include three reactions --
pp, $^7$Be, and $^8$B. The quantity $p_j^{ex}$ is the probability that 
in a particular experiment the neutrino arose from nuclear reaction $j$. We take
these from the analysis of Ref.~\cite{neutreview} for the solar experiments: 
chlorine \cite{homestake},
gallium (Sage,Gallex, and GNO) \cite{gallium}, SNO
\cite{sno}, and SNO-salt \cite{snosalt}. The function $f_j(r)$ 
is the probability that a neutrino is created by reaction $j$ at a radius $r$ 
\cite{bp2000} of the sun and is integrated from the center of the sun to the 
solar radius. The function
$g_j(E)$ is the energy distribution of the neutrinos emitted in reaction $j$. 
For $^7$Be this is a delta function at 0.88 MeV; the lower emission line does 
not contribute significantly. For the pp neutrinos, the energy distribution 
times the detector acceptance is a relatively narrow function of energy;
we set $E$ to its average. For $^8$B neutrinos, we 
use the energy distribution from the standard solar model \cite{bp2000}
and numerically perform the integration. 

For three neutrino mixing, the energy dependence of the solar data is well reproduced
by the MSW effect without level crossing. This is true of all the parameter sets examined 
here. The adiabatic approximation is thus justified after the fact.

The reactor experiments that we include are CHOOZ 
\cite{chooz} and KamLAND \cite{kamland}.  KamLAND is unique among 
reactor experiments as it 
measures ${\cal P}_{ee}$ where its predecessors set limits on $1-{\cal P}_{ee}$. It also
provides the energy spectrum of the neutrinos which tightly constrains the small mass squared 
difference. We list the value of ${\cal P}_{ee}$ in the table, but we actually fit with our
model the energy spectrum. In order to incorporate the systematic error for KamLAND, we introduce 
a normalization of their data, $N_{Kam}$, and float it constrained by an error of six percent. 
We also include the K2K experiment \cite{k2k} that measures the survival of muon neutrinos 
${\cal P}_{\mu\mu}$ over a long baseline (250 km) from KEK to the Super-K detector. 
The experiment, quantity measured, the value of that quantity to which we fit, and the 
average value of $L/E$ for each experiment are given in Table~\ref{t1}.

\begin{table}
\begin{center}
\begin{tabular}{|l|c|c|c|}  
\hline\hline
~Experiment~ & Measured & $L/E$ (m/MeV) & Data \\ \hline\hline
~Chlorine& ${\cal P}_{ee} $ & $4.0 \times 10^{10}$ &~$.337 \pm .065 $~ \\ \hline
~Gallium & ${\cal P}_{ee} $ & $35. \times 10^{10}$ &$.550 \pm .048 $ \\ \hline
~SNO     & ${\cal P}_{ee} $ & $2.2 \times 10^{10}$ &$.348 \pm .073 $ \\ \hline
~SNO-salt& ${\cal P}_{ee} $ & $2.2 \times 10^{10}$ &$.306 \pm .035 $ \\ \hline
~CHOOZ   & ${\cal P}_{ee} $ & 300. &$> 0.96$ \\ \hline
~KamLAND & ${\cal P}_{ee} $ & $4.1 \times 10^4$ &$.686 \pm .006 $ \\ \hline
~K2K     & ${\cal P}_{\mu\mu} $ & 208.            & $.55 \pm .19 $\\ \hline
\end{tabular}
\end{center}
\caption{The experiment, quantity measured, the average value of $L/E$, and 
experimental data for those quantities fit by the model are presented. We have combined the
SuperK and SNO results into one data point listed as SNO. We list 
${\cal P}_{ee} $ for KamLAND but we in fact fit the measured energy spectrum.}
\label{t1}
\end{table}

\begin{figure}
\includegraphics[width=3in]{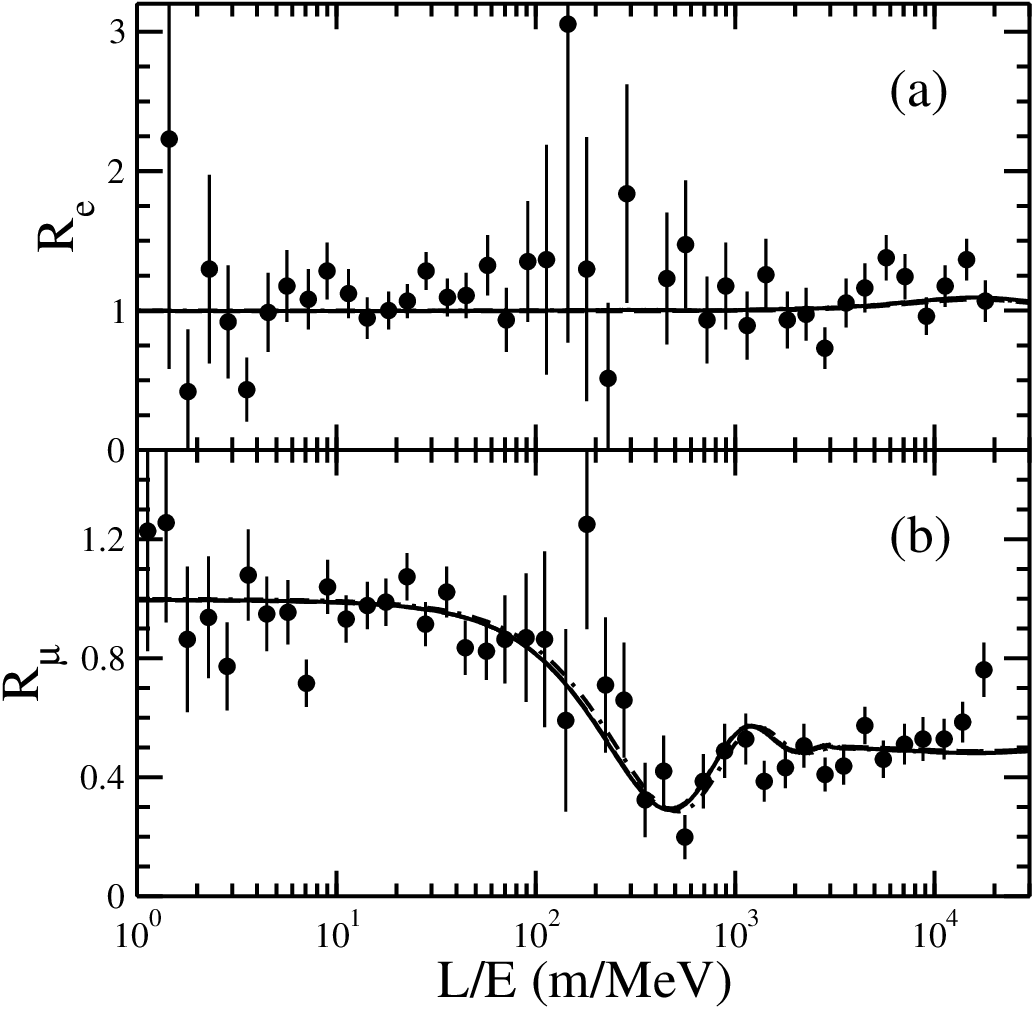}
\caption{Data and fits for the Super-K atmospheric experiment:  (a) electron-like
and (b) muon-like events.  Data is represented by points.  Fit 1 and fit 2, which 
is not distinguishable, yield the solid line; 
solution SS, the dashed-dotted line, is only barely distinguishable.} 
\label{fig1}
\end{figure}

The Super-Kamiokande experiment \cite{atm} has measured neutrinos that
originate from cosmic rays hitting the upper atmosphere. The detector 
distinguishes between $e$-like (electron and anti-electron) neutrinos and 
$\mu$-like (muon and anti-muon) neutrinos. The rate of $e$-like
neutrinos of energy $E$ arriving at the detector from a source a distance $L$ 
away is
\begin{equation}
{\cal R}_e (L,E) = {\cal P}_{ee}(L,E) + n(E) {\cal P}_{e\mu}(L,E)\,,
\end{equation}
and for $\mu$-like neutrinos
\begin{equation}
{\cal R}_\mu (L,E) = {\cal P}_{\mu \mu}(L,E) + \frac{1}{n(E)} {\cal P}_{e\mu}(L,E)\,,
\end{equation}
where $n(E)$ is the ratio of $\mu$-like neutrinos to $e$-like neutrinos at the 
source. We incorporate the Super-K atmospheric neutrinos by utilizing 
the $L/E$-dependence of ${\cal R}_e$ and ${\cal R}_\mu$ given in \cite{atm} and pictured in 
Fig.~\ref{fig1}. We take 
$n(E)$ to be 
energy independent and equal to 2.15.  As the absolute flux of cosmic rays striking the atmosphere is
not known to within $15 \%$, we introduce as a fit parameter 
an energy-independent renormalization factor $N_{atm}$ that multiplies the experimental 
values of $\mathcal{R}_e$ and $\mathcal{R}_\mu$.

The ratios ${\cal R}_e$ and ${\cal R}_\mu$ are convenient for the theorist as these are easily 
calculable. A distinct advantage of the atmospheric data is that for 
the neutrinos arriving from directly overhead to those arriving from the opposite
side of the Earth, the value of $L/E$ changes by almost four orders of magnitude. This is 
the only data which varies $L/E$. On the other hand, the source of neutrinos from
cosmic rays hitting the atmosphere must be modeled. Also,  the relationship between the 
direction of the recoil electrons in the detector and the direction of the neutrino initiating 
the reaction requires additional modeling. Thus the connection between the quantity 
measured and a simple physical parameterization is indirect and difficult to incorporate.
The details of the model can be found in \cite{model}.

\begin{table}
\begin{center}
\begin{tabular}{|c|c|c|c|} \hline\hline
parameter$\backslash$fit& SS & \#1 & \#2  \\ \hline\hline
$\chi^2_{dof}$ & ---  & ~1.19~ & ~1.21~  \\ \hline
$\theta_{12}$ &~$0.54\pm 0.05$~~& 0.48 & 0.49    \\ \hline
$\theta_{13}$~ & $\le 0.13$  & 0.12 & $-.04$   \\ \hline
$\theta_{23}$ & $0.79\pm0.06$  & 0.81 & 0.71  \\ \hline
$\Delta m^2_{21} \times 10^{-5}({\rm eV})^2$ &$8.3\pm 0.4$ & 7.6 &7.7 \\ \hline
$\Delta m^2_{32} \times 10^{-3}({\rm eV})^2$ &$2.4\pm 0.3$& 2.6 & 2.6 \\ \hline
$N_{atm}$        & --- & 1.00 & 1.00                            \\ \hline
$N_{Kam}$        & --- & 1.00 & 1.00                            \\ \hline
\end{tabular}
\end{center}
\caption{The value of $\chi^2_{dof}$ and the parameters for the standard solution 
(SS) and for the two local minima found here.}
\label{t2}
\end{table}

\begin{figure}
\includegraphics[width=3.in]{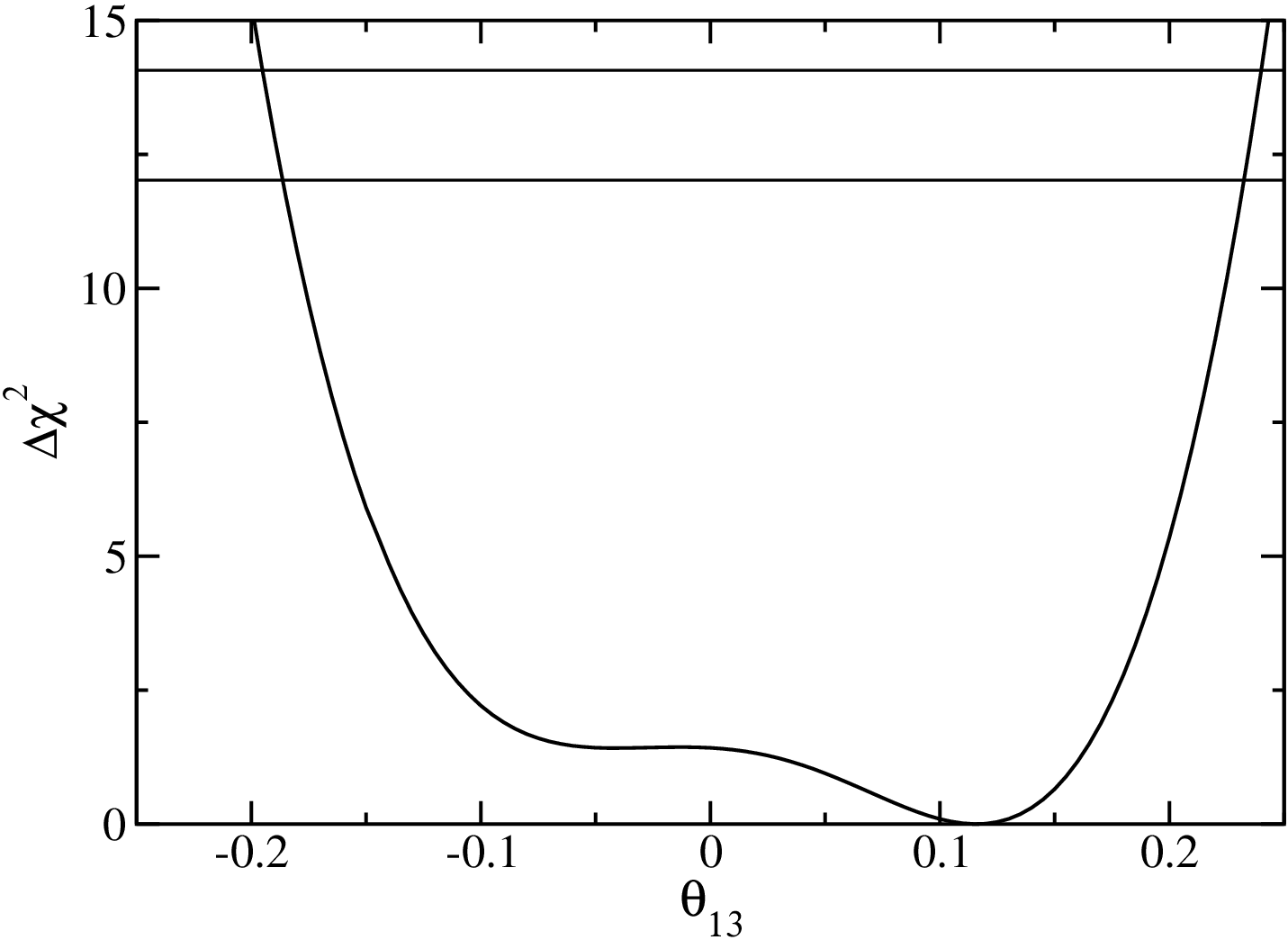}
\caption{The value of $\Delta \chi^2$ for a fixed value of $\theta_{13}$ with all other parameters 
varied. The horizontal lines indicate the 90\% and 95\% CL.}
\label{fig2}
\end{figure}

We fit the mixing angles, the mass squared differences, $N_{atm}$ and $N_{Kam}$ to the quantities in
Table~\ref{t1} and to the $L/E$ dependence of ${\cal R}_e$ and ${\cal R}_\mu$ pictured in 
Fig.~\ref{fig1} by minimizing chi-squared per degree of freedom, $\chi^2_{dof}$. In Fig.~\ref{fig2} we present
$\Delta \chi^2 =: \chi^2_{dof}-\chi^2_{min}$ as a function of $\theta_{13}$, where for each value of $\theta_{13}$
we have minimized with respect to the other parameters. The results are {\it not} symmetric about $\theta_{13}=0$, and we find 
two minima. The absolute minimum is at $\theta_{13}=0.12$, and there is a second local minimum at $\theta_{13}=-0.04$.
This asymmetry and the existence of the second local minimum would not exist if the commonly used factorization
approximation to the oscillation probabilities were employed, as this gives oscillation probabilities that
are functions of $\sin^2\theta_{13}$.

In Table~\ref{t2} we present the oscillation parameters for the two minima which we have found and also for a solution where 
$\Delta m^2_{21}$, $\theta_{12}$, and $\theta_{13}$ are taken from the analysis of Ref.~\cite{bahc}, and 
$\Delta m^2_{32}$ and $\theta_{23}$ are taken from Ref.~\cite{gonz}. We term this latter solution the 
``standard solution" (SS). We see that the parameters we find, particularly for the absolute minimum with
$\theta_{13}>0$, are reasonably consistent with those for the standard solution. We remind the reader that
we built the model \cite{model} not to extract precise values of the oscillation parameters, but to examine 
features of neutrino oscillation phenomenology in a semi-quantitative way. Here, we use the model to investigate 
the role of the negative $\theta_{13}$ region.

\begin{table}
\begin{center}
\begin{tabular}{|c|c|c|c|c|c|} \hline\hline
Experiment& Data & SS & \#1 & \#2 \\ \hline\hline
Chlorine& ~$.337\pm .065$~ & ~.451~ & ~.448~ & ~.454~ \\ \hline
Gallium & $.550\pm .048$ & .578 & .615 & .623\\ \hline
SNO     & $.348\pm .075$ & .395 & .371 & .378\\ \hline
SNO-salt& $.306\pm .035$ & .395 & .371 & .378\\ \hline
CHOOZ   & $>.96$         & .98 & .96 & .99\\ \hline
KamLAND & $.686\pm .006$ & .577 & .661 & .670\\ \hline
K2K     & $.55\pm .19$ & .60 & .56 & .57\\ \hline
\end{tabular}
\end{center}
\caption{The experimental results and the predictions for each from the models
whose parameters are given in Table~\protect\ref{t2}.}
\label{t3}
\end{table}

\begin{figure}
\includegraphics[width=3in]{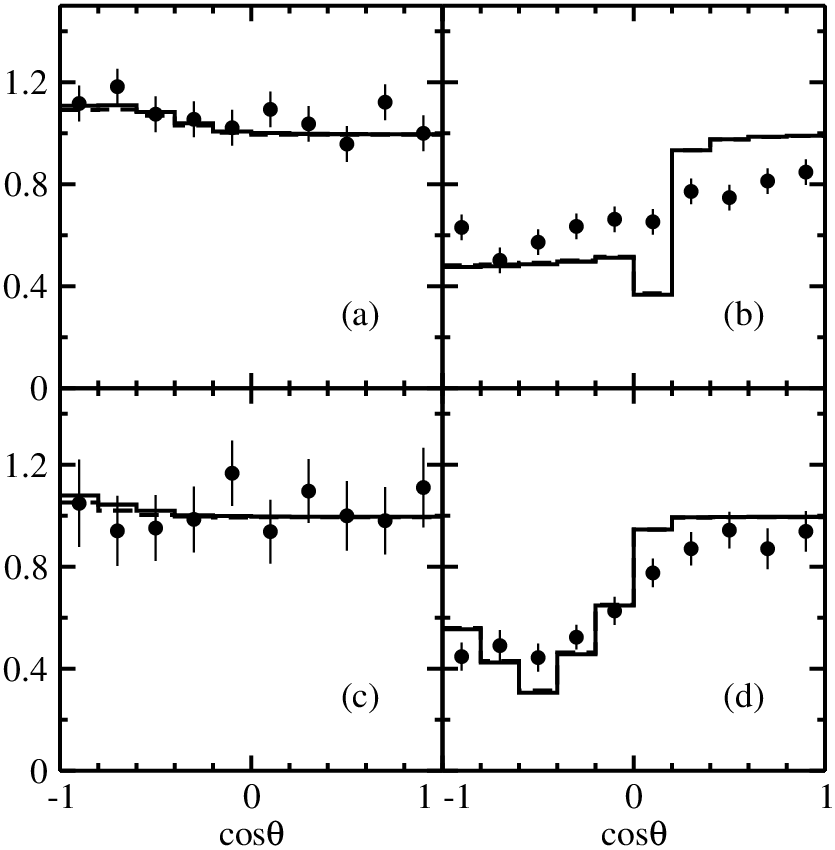}
\caption{Zenith-angle distributions for atmospheric Super-K experiment.  Points represent the
data, and the solid line shows the results of Fit 1.  (a) Sub-GeV electron-like.  
(b)  Sub-GeV muon-like.  (c)  Multi-GeV electron-like.  (d)  Multi-GeV muon-like and PC events. 
\label{fig3}}
\end{figure}

In Table~\ref{t3} we compare the data with the results of the fits, and in 
Fig.~\ref{fig1} we depict the $L/E$ dependence of each fit 
as compared to the atmospheric data.  In Fig.~\ref{fig1} curves are drawn 
for all three solutions. However, the results are sufficiently similar that the
individual curves cannot be distinguished. The model treatment of the atmospheric data is thus seen
to be quite comparable to a full analysis. The resulting fits to the data in Table~\ref{t3} are also seen to be reasonable. 

In order to further demonstrate that our results are reasonable, we calculate 
the zenith-angle dependence of the atmospheric data.  Using the energies defined for the various
classes of neutrino events in \cite{atm}, we determine $\mathcal{R}_e$ and $\mathcal{R}_\mu$ for 10
bins ranging from downward going ($\cos{\theta}=1$) to upward going ($\cos{\theta}=-1$) neutrinos.
We also allow for a simple, but more realistic, energy dependence for $n(E)$, taken from \cite{honda};
additionally, we introduce some overlap of the bins.  
We compare our results for the azimuthal dependence of the neutrinos to the dependence of 
the observed recoil electrons  
seen at Super-K, normalized to their no-oscillation Monte Carlo
simulation, in Fig.~\ref{fig3}.  Though we do not model the recoil electron, there is a
strong correlation between the two processes for the high-energy events.  The results
are encouraging.
For the lower energies,
all the solutions produce little electron neutrino oscillations as indicated by the data.
However, there is a visible low-energy muon neutrino oscillation which is larger in the theory 
than in the data.  An improved model
of the atmospheric data is required to better understand this. Most importantly, 
the high-energy muon zenith-angle data is qualitatively similar to the results given by 
our model.

In summary, within the model developed in Ref.~\cite{model} of the neutrino oscillation data, we find 
that the region $\theta_{13} < 0$ plays an important role in understanding
the oscillation parameters for three-neutrino oscillations. As oscillation probabilities
for negative and positive values of $\theta_{13}$ are not simply related, the analysis
cannot be performed in terms of $\sin^2 \theta_{13}$. The work presented here is intended to 
motivate a more thorough and careful examination of the $\theta_{13} < 0$ region of 
the parameter space.

\begin{acknowledgments}
The authors are grateful for very helpful conversations with D. V. Ahluwalia and
I. Stancu. This work is supported by the U.S. Department of Energy under 
grant No.~DE-FG02-96ER40963.
\end{acknowledgments}

\bibliographystyle{unsrt}
\bibliography{nolsnd2}
\end{document}